\documentclass[runningheads]{llncs}

\usepackage[T1]{fontenc}
\usepackage{graphicx}
\usepackage{hyperref}

\usepackage{booktabs}

\usepackage{todonotes}
\begin{document}
\title{Enhancing Information Retrieval in Digital Libraries through Unit Harmonisation in Scholarly Knowledge Graphs}
\titlerunning{Unit Harmonisation in Scholarly Knowledge Graphs}
%
\author{Golsa Heidari \orcidID{0000-0002-5398-7086} \and
Markus Stocker \orcidID{0000-0001-5492-3212} \and
Sören Auer \orcidID{0000-0002-0698-2864}}
\authorrunning{G. Heidari, M. Stocker, and S. Auer}

\institute{TIB -- Leibniz Information Centre for Science and Technology \\ Hannover, Germany \\
\email{\{golsa.heidari, markus.stocker, auer\}@tib.eu} \\
}
\maketitle              
\begin{abstract}
Scientists have always used the studies and research of other researchers to achieve new objectives and perspectives. In particular, employing and operating the measured data in previous studies is so practical. Searching the content of other scientists' articles is a challenge that researchers have always struggled with. Nowadays, the use of knowledge graphs as a semantic database has helped a lot in saving and retrieving scholarly knowledge. Such technologies are crucial to upgrading traditional search systems to smart knowledge retrieval, which is crucial to getting the most relevant answers for a user query, especially in information and knowledge management. However, in most cases, only the metadata of a paper is searchable, and it is still cumbersome for scientists to have access to the content of the papers. In this paper, we present a novel method of faceted search \emph{structured content} for comparing and filtering measured data in scholarly knowledge graphs while different units of measurement are used in different studies. This search system proposes applicable units as facets to the user and would dynamically integrate content from further remote knowledge graphs to materialize the scholarly knowledge graph and achieve a higher order of exploration usability on scholarly content, which can be filtered to better satisfy the user's information needs. The state of the art is that, by using our faceted search system, users can not only search the contents of scientific articles, but also compare and filter heterogeneous data. 
\end{abstract}
\section{Introduction}
As digital libraries continue to evolve beyond their original role as repositories for publications, there is a growing need for systems that enable detailed, content-level retrieval across scholarly literature~\cite{alzahrani2024navigating}. While early digital libraries mainly focused on storing PDFs and indexing bibliographic metadata, today’s research environment demands more advanced and interactive access to the semantic content of scientific articles~\cite{shah2025digital,davitaia2025recursive,khan2025traditional,oelen2024orkg}. Researchers not only need to find relevant documents, but also increasingly require access to the underlying data, experimental results, and empirical findings presented within those documents. These elements—such as physical measurements, performance benchmarks, or observational metrics—are critical for conducting meta-analyses, developing predictive models, and validating scientific hypotheses. However, despite advancements in search infrastructure and the growing application of AI in scientific research~\cite{oelen2024orkg,yu2025ai}, current digital library platforms still lack the ability to provide fine-grained access to detailed scientific content.  Traditional keyword-based search engines have difficulty processing structured data embedded within unstructured full-text articles~\cite{Khalid2021}. Although bibliographic metadata and citation graphs offer useful high-level summaries~\cite{kinney2023semantic}, they do not deliver the semantic depth needed for data-driven exploration or comparative analysis.

One of the most critical- and often overlooked—barriers in this regard is the heterogeneous representation of measured data~\cite{vahdati2016openresearch,vahdati2019collaborative}. Unlike metadata~\cite{hendricks2020crossref} (e.g., author, title, publication year), measured values are typically reported in diverse formats and with varying units of measurement, depending on the authors’ preferences, scientific domain, or regional conventions. This lack of standardization hinders efforts to compare, filter, or aggregate findings across studies. For instance, sea level rise might be reported as 0.25 meters in one study and 25 centimeters in another—semantically equivalent, but syntactically distinct. Without automated unit normalization and semantic alignment, such inconsistencies pose significant challenges for information retrieval systems, making it difficult to ensure comprehensive and accurate search results across large corpora of scientific texts. To address these challenges, digital libraries must move beyond surface-level indexing and adopt content-aware, semantically enriched retrieval mechanisms. One promising direction is the use of Scholarly Knowledge Graphs (SKGs)~\cite{jaradeh2019open,vogt2020toward,verma2023scholarly}—structured graph representations of scholarly entities, relationships, and concepts. SKGs provide a foundation for semantic search by modeling papers, authors, topics, and citations as interconnected nodes. They enable context-aware exploration and discovery, and have become an integral part of modern research infrastructures. However, despite their growing adoption, most existing SKGs focus primarily on bibliographic metadata, citation networks, and topic modelling, and offer limited support for quantitative content or unit-aware retrieval. As a result, the potential of SKGs remains underutilized in tasks that require the comparison of measured data across multiple sources. In this work, we build upon the Open Research Knowledge Graph (ORKG)~\cite{jaradeh2019open}, an open infrastructure designed to represent scholarly knowledge beyond traditional metadata semantically. ORKG provides structured descriptions of research contributions, methods, datasets, and findings, enabling machine-actionable access to scientific content at a fine-grained level. By leveraging ORKG as the underlying SKG, our framework inherits its interoperability, openness, and rich semantic modeling of scholarly artifacts. This foundation allows us to extend ORKG with unit-aware representations of measured data, thereby enriching its retrieval capabilities and supporting quantitative comparison across heterogeneous sources.

In this paper, we introduce a novel framework that enables \textit{unit-aware retrieval and comparison of measured data} in digital libraries using a structured, faceted search interface built on top of ORKG. Our system automatically extracts measured quantities and their associated units from the full text of scientific articles, normalizes these values to standard units using semantic mapping, and exposes them as interactive search facets. This allows users not only to query the content of scientific papers in a more targeted way, but also to filter and compare quantitative data across articles—even when originally reported using different units or formats. Such capabilities are especially valuable in interdisciplinary fields where unit conventions vary widely and where direct comparisons of experimental results are essential. Furthermore, our system is designed to be scalable and interoperable. It supports dynamic integration with remote and federated knowledge graphs, allowing researchers to query enriched scholarly content across multiple sources. This extensibility enhances both the breadth of available knowledge and the depth of semantic exploration, enabling more meaningful interactions with scientific data through digital library platforms. By leveraging structured knowledge representations and unit harmonization techniques, we offer a retrieval framework that supports not just document discovery but also scientific reasoning and comparative analysis. Overall, this work aims to contribute to the ongoing transformation of digital libraries into intelligent research environments—ones that go beyond document storage to actively support exploration, comparison, and reuse of scientific knowledge. By harmonizing measured data and enabling unit-aware interaction, we help bridge the gap between traditional bibliographic search and content-driven scientific discovery.

The remainder of the paper is organized as follows: \autoref{related-work} reviews related work on information extraction and semantic search; \autoref{methodology} details the methodology and system architecture; \autoref{implementation} presents implementation and integration aspects; \autoref{discussion} discusses results, challenges, and broader impact; and \autoref{conc-future-work} concludes with future research directions.

\section{Related Work}
\label{related-work}

Knowledge graphs, the representation of information as a semantic graph, have caused broad interest in both the industrial and academic worlds. Their nature of providing semantically structured information has brought substantial possible solutions for many tasks, including question answering, recommendation, and information retrieval, and is considered to offer great promise for enabling more intelligent machines by numerous researchers~\cite{Zou_2020,oelen2021smartreviews}. As a flexible and expressive data model, knowledge graphs are increasingly used in digital library systems to represent and interlink diverse scholarly entities—such as authors, publications, institutions, and research topics—in a unified semantic structure. One of the most important features of a knowledge graph as a database is data extraction, which enables automatic population and enrichment of the graph from unstructured or semi-structured sources. In the context of scholarly content, this includes extracting meaningful scientific information such as entities, relations, and measured data from full-text documents.

Given the scope of this paper, the related work is organized into two main parts. First, we review relevant literature on Information Extraction in Knowledge Graphs~\cite{jaradeh2019open,jaradeh2020question,jaradeh2021plumber,brack2021citation,tzitzikas2017faceted,sanchez2020lindasearch}, with a particular focus on techniques for identifying and normalizing measured data from scholarly texts. Second, we examine prior work on Faceted Search in Knowledge Graphs~\cite{auer2020improving,stocker2023fair,haris2022enriching,heidari2021leveraging,heidari2021demonstration,arenas2016faceted,papadaki2024unifying,rantala2024representing,meesad2024knowledge}, which explores how semantic facets derived from structured data can support rich, interactive exploration in digital libraries. This part of the review highlights the role of faceted interfaces in improving usability and discovery, especially in scenarios involving heterogeneous or quantitative data. 

\subsection{Information Extraction in Knowledge Graphs}
Scientists engage in research and development efforts aimed at enhancing document-based scholarly communication by incorporating semantic representations of scholarly knowledge. They have proposed an infrastructure architecture along with several of its core features, which enable the acquisition of scholarly knowledge in a machine-actionable format. This, in turn, supports the processes of scholarly knowledge curation, publication, and retrieval. They employed a combination of crowdsourcing and automated natural language processing (NLP) techniques, thereby establishing a multi-modal scholarly knowledge acquisition approach designed to improve the search and exploration of scholarly content~\cite{jaradeh2019open}.

JarvisQA~\cite{jaradeh2020question} is a BERT-based transformer system designed to answer questions about tabular views of scholarly knowledge graphs. It can generate direct responses to a wide range of natural language questions (NLQs) related to tabular data found in scientific articles. The authors also introduced a preliminary dataset comprising relevant tables and corresponding NLQs. In a related effort, Plumber~\cite{jaradeh2021plumber} is a modular framework that supports both manual and automated construction of information extraction pipelines. Jaradeh et al. further proposed a method for extracting triples from text and aligning them with existing knowledge graphs.

Citation recommendation in research articles plays a crucial role in assisting researchers to enhance the quality of their work by suggesting relevant and related literature. Brack et al.~\cite{brack2021citation} proposed leveraging research knowledge graphs, which interconnect research papers through shared scientific concepts, as a means to support this task. Their study investigated the effectiveness of an automatically populated research knowledge graph in improving citation recommendations. To achieve this, they integrated document embeddings—learned from both the textual content and the citation network—with concept vectors that represent scientific concepts.

Y. Tzitzikas et al.~\cite{tzitzikas2017faceted} presented an analysis of features and standards essential for browsing and exploring RDF/S datasets. They identified key information requirements and structural components involved in such systems. Their work introduced a generalized model to unify the primary approaches to faceted exploration and browsing, which is based on a small set of abstract states and transitions between them. Similarly, the LINDASearch system~\cite{sanchez2020lindasearch} provides a middleware platform designed to deliver unified access to information across various Open Linked Data projects. These include datasets such as DBpedia, GeoNames, LinkedGeoData, FOAF profiles, the Global Health Observatory, the Linked Movie Database (LinkedMDB), and World Bank Linked Data.

\subsection{Faceted Search in Knowledge Graphs}
Although faceted search is extremely beneficial for information retrieval, search engines have used it almost at the level of metadata for the scholarly literature. In some research fields, researchers also depicted content in articles in a structured manner, and they have built search systems, which, of course, is helpful. But their work is limited to one discipline.

The Open Research Knowledge Graph (ORKG)~\cite{auer2020improving,stocker2023fair} is a service that represents information in a knowledge graph, which is readable by machines and humans. Its comparison faceted search is reached by introducing contextual facets (e.g., citations). The developed approach enables presenting ORKG scholarly knowledge flexibly improved with contextual information sourced in a federated manner from numerous technologically heterogeneous scholarly communication infrastructures~\cite{haris2022enriching}. They proposed a web widget-based approach for dynamic retrieval and display of comprehensive contextual information for scholarly knowledge. The approach enables information presentation and is powered by a GraphQL-based federated query service that virtually integrates and abstracts the technological heterogeneity of scholarly communication infrastructures.

Faceted search has become a prominent method for exploring structured scholarly content through semantic technologies. Heidari et al.~\cite{heidari2021leveraging,heidari2021demonstration} propose a methodology for enhancing a faceted search system by leveraging a federation of scholarly knowledge graphs. Their approach is implemented on top of a scholarly knowledge graph and supports integration with third-party knowledge graphs, improving the breadth and depth of scholarly content exploration.

Building on the theoretical foundations of faceted search, Arenas et al.~\cite{arenas2016faceted} offer formal models within the context of RDF-based knowledge graphs enriched with OWL ontologies. In addition to theoretical contributions, other researchers have proposed practical extensions, such as facet ranking systems and mechanisms for filtering large answer sets by applying statistical constraints to reduce information overload. Expanding on these concepts, Papadaki and Tzitzikas~\cite{papadaki2024unifying} introduced a unified framework that combines faceted search with analytical operations over RDF knowledge graphs. Their system supports both filtering and interactive analytics, allowing users to explore datasets more dynamically and semantically. Similarly, Rantala et al.~\cite{rantala2024representing} developed a faceted relational search framework specifically tailored for cultural heritage knowledge graphs. Their system enables the exploration of entity relationships—such as between people and places—through dynamically generated facets and visualizations that highlight relational patterns and aggregated statistics.

In a broader context, Meesad and Mingkhwan~\cite{meesad2024knowledge} offer a comprehensive review of knowledge graphs within smart digital libraries. They examine key implementation strategies and real-world applications, particularly emphasizing how knowledge graphs enhance data integration, discoverability, and personalized user experiences in digital library environments.

\section{Methodology}
\label{methodology}
In faceted search systems, units can be presented to users as selectable facets. The core idea is to utilize measured data from scientific studies and enable dynamic unit conversion based on the user’s selected unit, making the data directly comparable. This allows users to filter the data more effectively and identify results that precisely match their queries. To support this functionality, the underlying scholarly knowledge graph must include semantic, machine-readable representations of the measured data. This knowledge graph should capture structured elements of a publication, interlinking properties such as the measured numeric value, the quantity kind, the unit, and a standardized unit code compatible with the unit conversion service. In the following sections, we describe the conceptual foundation and the step-by-step workflow of the proposed methodology.

\subsection{Conceptual Model}

\begin{figure}[t]
  \centering
  \includegraphics[width=\linewidth]{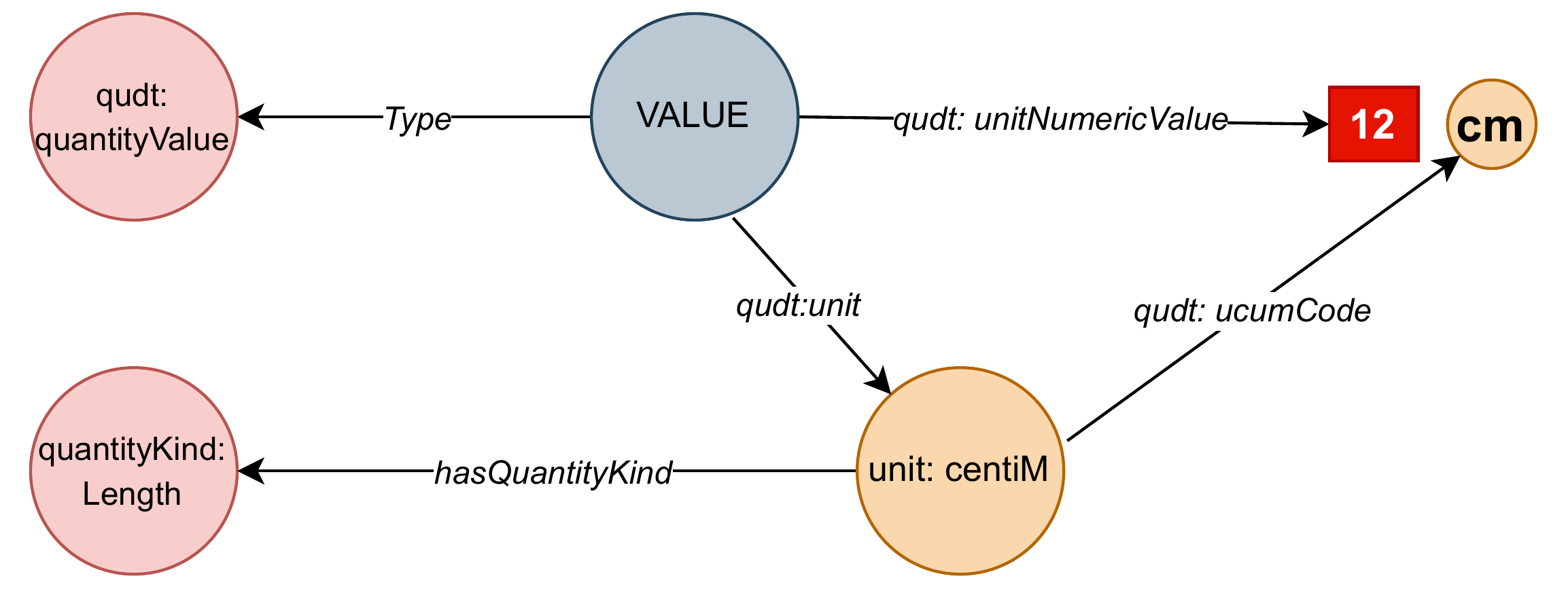}
  \caption{Conceptual model for representing measured data in ORKG, aligned with the QUDT ontology. Each measurement is modeled as a \texttt{quantityValue} entity that links a \texttt{numericValue}, a \texttt{unit} (with its UCUM code), and a corresponding \texttt{quantityKind}. This structure enables semantic comparability, unit-aware filtering, and support for dynamic unit conversion in the faceted search system.}
  \label{conceptual}
\end{figure}

Scientific studies often report measurements using different units. For instance, one study might state "The tumour size was recorded as 0.12 metres", while another reports "The tumour size was recorded as 12 centimetres". If a user searches using the unit "meters", the second study may be excluded from results or ranked lower, even though both values are equivalent. To overcome such inconsistencies, our search system moves beyond simple keyword matching. It standardizes and semantically aligns measured data across different units, ensuring that equivalent information can be retrieved uniformly. A key concept in our approach is the classification of each measurement by its underlying quantity kind (e.g., length, mass, temperature), which allows for meaningful comparisons and filtering. This motivates the need for a structured, ontology-driven representation of measured data, which we formalize in the following conceptual model.

As we showed in \autoref{conceptual}, when a user initiates a query with a specific unit (e.g., "12 cm"), the system automatically maps this request to its underlying quantity kind (e.g., Length) and triggers a unit conversion service. This service converts the original data from the publication into the unit selected by the user. As a result, the data becomes semantically and numerically comparable, enabling meaningful filtering across different unit expressions. This unit-aware transformation supports flexible faceted search functionality. Users can search not only for exact values but also filter results using range-based constraints (e.g., values greater than, less than, or within a certain interval). Additionally, exclusion filters can be applied (e.g., exclude values in a given range). The system also offers an auto-completion feature that proposes valid and context-appropriate units based on the selected quantity kind, improving usability and reducing input errors.

To implement this conceptual infrastructure, we leveraged the QUDT (Quantity, Unit, Dimension, and Type) ontology~\cite{fairsharing_qudt_2025}. QUDT provides a formal framework for representing quantities, quantity kinds, units, dimensions, and their interrelations. Its compliance with international standards ensures interoperability and semantic consistency across systems. Next, we designed our SKG~\footnote{This work utilizes ORKG (accessible via~\url{https://orkg.org/}) as its underlying infrastructure.} in alignment with the QUDT ontology design pattern. As illustrated in \autoref{conceptual}, our model interlinks the following components:
\begin{itemize}
    \item \texttt{quantityKind}: The abstract physical concept being measured (e.g., \texttt{Length}).
    \item \texttt{unit}: The measurement unit (e.g., centimetre, with \texttt{qudt:ucumCode = 'cm'}).
    \item \texttt{Value}: The numerical value associated with the measurement (e.g., 12).
    \item \texttt{quantityValue}: A composite entity that binds the numeric value to the unit and quantity kind, forming a semantically rich representation of the measured data.
\end{itemize}
This structured representation allows us to semantically reason over measured data and facilitate advanced, unit-aware search interactions within scientific content.

\subsection{Workflow}

\begin{figure*}[t]
  \centering
  \includegraphics[width=\textwidth]{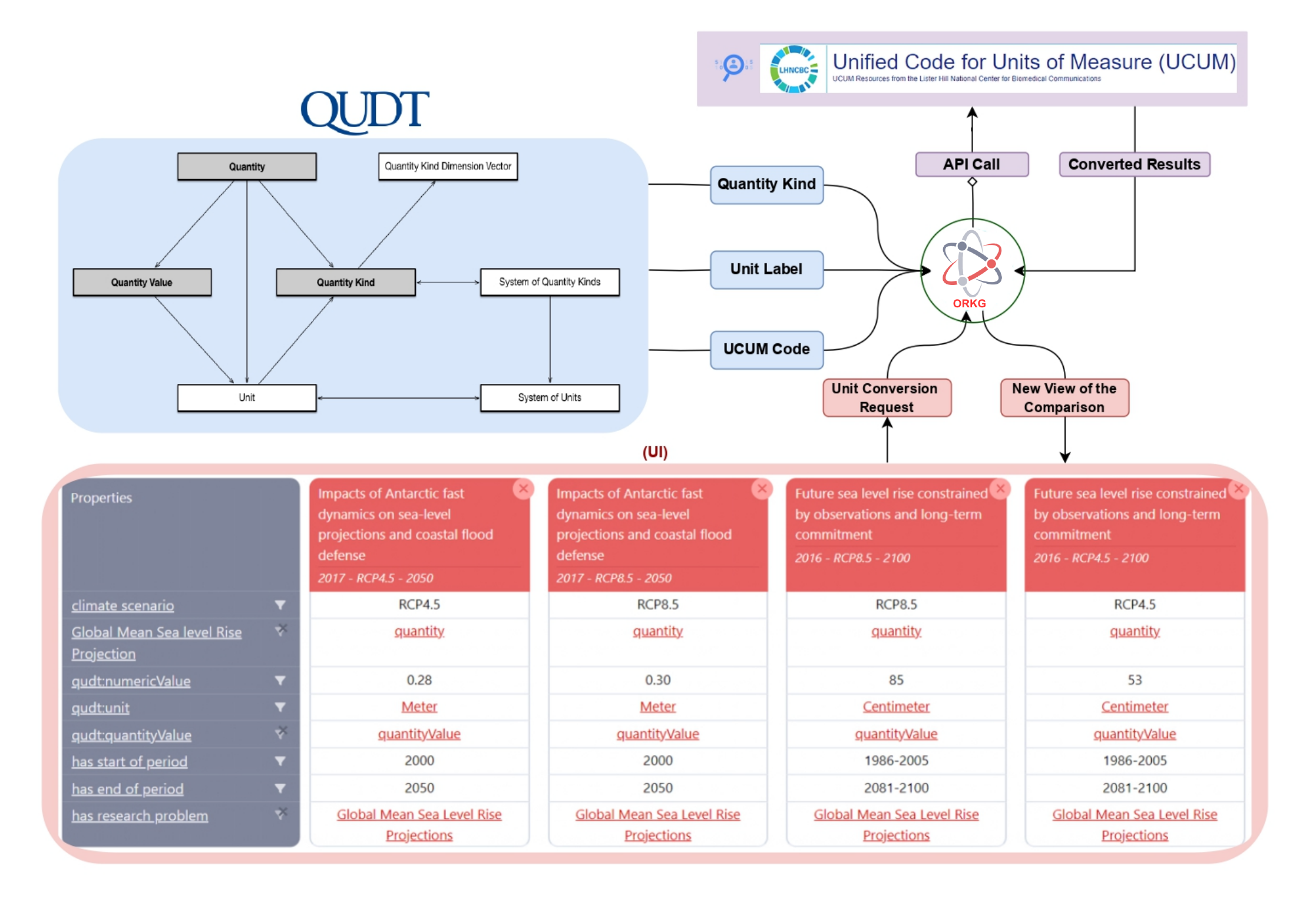}
  \caption{Workflow of unit harmonization using ORKG scholarly knowledge graph, QUDT ontology, and UCUM-based unit conversion service. The system processes user-selected units, materializes metadata from QUDT, performs conversions via UCUM, and presents harmonized results for comparison.}
  \label{workflow}
\end{figure*}

The system workflow consists of three important components, such as QUIDT ontology,  UCUM (Unified Code for Units of Measures) service~\footnote{\url{https://ucum.nlm.nih.gov/}}~\cite{schadow2005unified}, and ORKG in the center that responds to unit conversion requests made in the user interface (UI). The \autoref{workflow} illustrates the interaction and how measured data is processed between ORKG, the QUDT ontology, the UCUM-based unit conversion service, and the UI. The process involves the following steps:

\begin{enumerate}
    \item \textbf{User Interaction}:  The UI allows end-users to explore measured data and interactively select a preferred target unit for comparison, filtering, or harmonization. This selection can be made from a dynamically generated comparison table showing different unit representations linked to the underlying data sources.
    
    \item \textbf{Unit Conversion Request}: Once the user selects a desired unit from the comparison table, a formal \textit{Unit Conversion Request} is triggered, where ORKG sends a request to an external UCUM-compliant web service via an \textit{API Call}. This request contains the original measurement value along with its source unit and the desired target unit, both referenced using their respective UCUM codes. This standardization enables precise and reliable communication with the conversion backend.
    
    \item \textbf{Units Materialization}:  Upon user selection, ORKG performs semantic inference to \textit{materialize} unit-related facets about the chosen unit before the API call to UCUM. Materialization involves retrieving the associated \textit{Quantity Kind}, \textit{Unit Label}, and \textit{UCUM Code} from the QUDT-encoded knowledge graph. ORKG ensures that all semantically valid information about the unit is made explicit and ready for conversion.
    
    \item \textbf{Units Conversion using UCUM}: The UCUM is a system for coding units of measure that was developed with the aim of representing all units of measurement used internationally in science, including laboratory medicine and pharmaceuticals. Using UCUM service processes the request and returns a response containing the \textit{converted value}. This result is expressed in the target unit as specified by the user and computed using reliable and standardized conversion logic defined by the UCUM specification.
    
    \item \textbf{Data Presentation}:  Finally, the system presents a harmonized view of the data to the user, a \textit{New View of the Comparison}. The original values—possibly coming from multiple heterogeneous sources—are now converted into a consistent unit, enabling direct and meaningful comparison. The UI updates to reflect the transformed data, offering better interpretability and enhanced usability for data-driven tasks.
\end{enumerate}

This workflow ensures that measured data from various sources can be harmonized and compared seamlessly, despite differences in the original units used in the publications.

\section{Implementation}
\label{implementation}

\begin{figure*}[t]
  \centering
  \includegraphics[width=\textwidth]{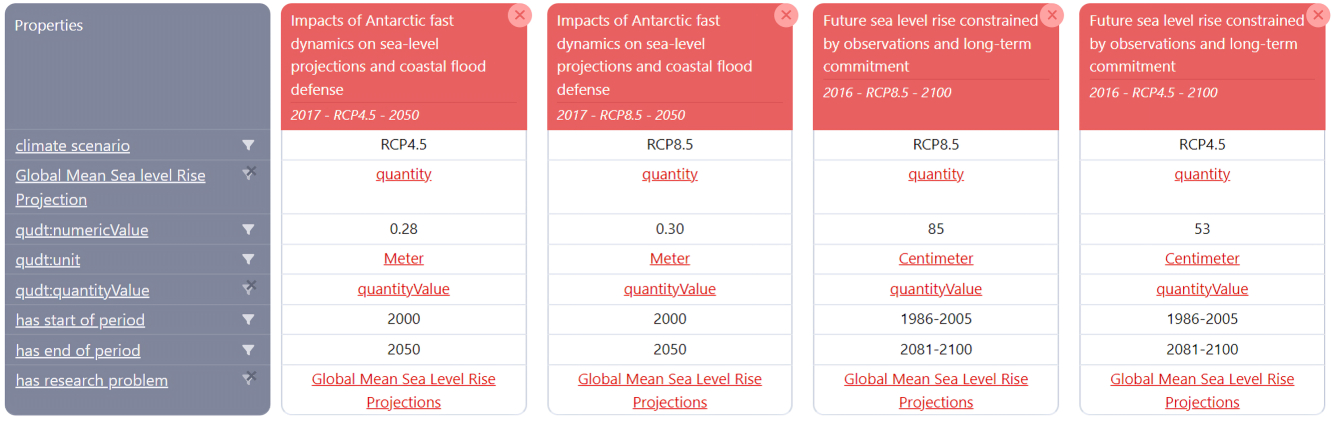}
  \caption{Illustration of ORKG comparison, which enables a unit conversion.  The comparison is available at \url{https://orkg.org/comparisons/R175109}.
  }
  \label{Comparison-table}
\end{figure*}

ORKG, as a building block in our unit harmonization workflow, is an online resource that semantically encodes research contributions as an interconnected knowledge graph~\cite{oelen2020creating}. It transforms traditionally prose-based scholarly literature into a machine-readable format~\cite{fathalla2017towards}, enabling structured access to research content. ORKG also supports the publication of tabular representations of contributions, allowing for systematic comparisons. Given a set of papers and their associated research contributions, ORKG facilitates the comparison of how different works address a common research problem across the literature. The \autoref{Comparison-table} depicts a comparison in ORKG, which compares different \textit{global mean sea level rise projections} measured in various studies. We implemented our search system for ORKG comparisons, where research contribution descriptions are specified by predefined templates. These templates support the unit conversion in ORKG comparisons. 

The QUDT ontology offers a variety of relations for the described resources, and according to its schema, each instance of the unit class in ORKG has a link to the corresponding resource in the QUDT ontology. In which we are interested in the \textit{UCUM Code}\footnote{Unified Code for Units of Measure}, which is used by the UCUM service\footnote{\url{https://ucum.nlm.nih.gov/ucum-service.html}} for converting units that have the same quantity kind.

To illustrate how our knowledge graphs support unit conversion, consider the example of the \textit{global mean sea level} property in \autoref{Comparison-table}. Its associated \textit{quantity} template includes two key components: \textit{has quantity kind}, which identifies the type of measurement (e.g., Meter) and determines the applicable units that can be shown to users for conversion, and \textit{quantity value}, which holds the actual data. The \textit{quantity value} is further structured into two elements: the \textit{numeric value} and the associated \textit{unit}, where each unit is an instance of the unit class. For every unit instance, we retrieve its \textit{quantity kind}, label, and \textit{UCUM code} using the QUDT SPARQL endpoint~\footnote{\url{https://www.qudt.org/fuseki/}}. This information allows us to define a reusable unit template, which we then materialize in ORKG. The extracted UCUM code is used to make API calls to the UCUM conversion service, following this request format: \texttt{\{source\_quantity\}/from/\{source\_unit\}/to/\{target\_unit\}}.  Where the 
\texttt{source\_unit} refers to the unit extracted from the content of a paper, while the \texttt{target\_unit} is chosen by the user from dynamically generated facets. For both units, their corresponding UCUM codes are used to ensure standardized conversion. We implement the solution by sending an API request to the UCUM service. The result of this API call is then presented to the user as a new, converted value.

The \autoref{Comparison-table} depicts an example of unit conversion in a \textit{Global Mean Sea Level Rise Projection} contribution comparison. When the calculator icon is clicked, a dialog box appears displaying valid unit options, allowing the user to select a preferred unit. When applying a calculator, the color of the icon changes to indicate that these values are not the main values of the contributions and are changed due to unit conversion. Additionally, a tooltip -- a small pop-up box that appears when a user hovers over the calculator icon-- about the selected unit is displayed.  The results are displayed on the screen in real time, but are not stored in the main database by default. However, the system allows users to save these configurations and the selected data subset as a new comparison entry, which is stored with a permanent, shareable URL for consistent access and comparison by other researchers.

\begin{table}[t]
    \centering
    \caption{Real-world example from ORKG contributions and the benefits for faceted search and filtering.}
    \resizebox{\textwidth}{!}{%
    \begin{tabular}{p{4.5cm}@{\hspace{0.5cm}}p{7cm} p{5cm}@{\hspace{0.5cm}}p{2.2cm}}
        \toprule
        \textbf{Original Measured Statement} & 
        \textbf{Structured Representation} & 
        \textbf{Search Benefit} & 
        \textbf{URL} \\
        \specialrule{1.5pt}{0pt}{0pt} 

        \textit{The BERT model achieved a micro F1 score of 85.6\% on the Natural Language Inference task.} &
        \begin{tabular}[t]{@{}l@{}}
        \texttt{quantityKind: micro F1 score} \\
        \texttt{unit: \%} (UCUM: \texttt{\%}) \\
        \texttt{value: 85.6} \\
        \texttt{quantityValue: \{micro F1 score, 85.6, \%\}}
        \end{tabular} &
        Enables filtering and comparison of model performances across benchmarks by micro F1 score &
        \textbf{Contribution-ID:} \href{https://orkg.org/papers/R756122}{R756122} \\
        \addlinespace[3mm] 

        \textit{The cut-off wind speed for the Polaris P20 wind turbine is 25 meters per second.} &
        \begin{tabular}[t]{@{}l@{}}
        \texttt{quantityKind: Speed} \\
        \texttt{unit: m/s} (UCUM: \texttt{m/s}) \\
        \texttt{value: 25} \\
        \texttt{quantityValue: \{Speed, 25, m/s\}}
        \end{tabular} &
        Enables filtering and comparison of wind turbines by cut-off wind speed across locations &
        \textbf{Contribution-ID:} \href{https://orkg.org/papers/R709075}{R709075}, \textbf{Comparison-ID:} \href{https://orkg.org/comparisons/R1092669}{R1092669} \\
        \addlinespace[3mm]

        \textit{The water content in steamed African carp (without salt) is 75.1 grams.} &
        \begin{tabular}[t]{@{}l@{}}
        \texttt{quantityKind: Mass} \\
        \texttt{unit: g} (UCUM: \texttt{g}) \\
        \texttt{value: 75.1} \\
        \texttt{quantityValue: \{Mass, 75.1, g\}}
        \end{tabular} &
        Enables nutritional comparison and filtering of food components by mass across different foods &
        \textbf{Contribution-ID:} \href{https://orkg.org/papers/R708567/R1363174}{R1363174} \\
        \addlinespace[3mm]

        \bottomrule
    \end{tabular}
    }
\label{tab:examples}
\end{table}

To further illustrate how our semantic knowledge graph approach benefits search and comparison across diverse scientific domains, ~\autoref{tab:examples} presents real-world examples of structured measured data from ORKG contributions. These examples demonstrate how representing quantities with their associated quantity kind, unit, and value enables flexible and accurate faceted search, filtering, and unit-aware comparison across datasets.

\section{Discussion}
\label{discussion}
Faceted search enhances the user experience in digital libraries by enabling users to filter search results and recover more relevant information. However, when it comes to scientific content, these benefits are often limited—particularly for researchers who require access to quantitative knowledge embedded in the content of publications, not just metadata. While many digital libraries and SKGs allow structured exploration, they fall short when dealing with measured data that is expressed in a variety of units. In such cases, the absence of intelligent unit-aware facets makes meaningful comparison across studies difficult or impossible. This section discusses the critical factors in evaluating and advancing search systems that aim to overcome these limitations.

\noindent \textbf{Precision Challenges.} A core quality metric of any retrieval system is precision—the extent to which retrieved results match the user's intent. Although many knowledge graphs discussed in prior works have impressive coverage of metadata and bibliographic information, they often lack deep structuring of the full-text content, especially when it comes to measured or numerical data. As a result, even if relevant articles exist, users are unable to search or filter based on specific data points, such as a measured rise in sea level or a concentration threshold in an experiment. Browsing semantically structured content rather than unstructured full text has the potential to improve precision significantly, but it also introduces new challenges, such as the need for more expressive query interfaces and support for complex user intents.

\noindent \textbf{Recall Limitations.} Equally important is recall, or the ability of the system to retrieve all relevant results. Some knowledge graphs do structure measured data, but their scope is often limited to a small subset of publications, domains, or datasets. Consequently, many potentially relevant results remain inaccessible to the user. A high-precision system with low recall still fails to satisfy the researcher's need for comprehensive knowledge discovery, especially in systematic reviews or comparative studies.

\noindent \textbf{Unit Harmonization Challenge.} Without converting all measured data to a common or user-specified unit, comparisons across papers lose their meaning. For example, a researcher examining studies on sea level rise may encounter results reported in meters, centimeters, or even inches. Our system addresses this by dynamically recognizing and converting units, ensuring that data across publications becomes directly comparable. This capability is integrated into our faceted search interface, which does not rely on predefined facets. Instead, it generates context-aware facets on the fly, depending on the nature of the query and the units found in the underlying data.

By offering dynamic unit-aware facets, our system enables researchers to narrow search results based on numeric value ranges in a standardized unit of their choosing. This not only improves precision but also enhances the usability and interpretability of the results. Furthermore, it bridges the gap between structured semantic search and content-level retrieval—something existing SKGs rarely offer in combination.

In summary, precision, recall, and unit harmonization are essential to effective search in scientific domains. Our proposed system contributes a significant advancement by enabling smart, dynamic filtering of structured measured data, making SKGs and digital libraries more powerful and practical tools for scientific research.

\section{Conclusions and Future Works}
\label{conc-future-work}
Effective retrieval of scientific knowledge goes beyond accessing bibliographic metadata—it requires the ability to understand, search, and compare the actual content of research papers. Measured data such as experimental values, physical quantities, and performance metrics are often central to scientific findings, yet remain difficult to search or compare due to variability in units and a lack of structured representation. While digital libraries and academic search systems have advanced significantly, most still rely on full-text indexing or surface-level metadata, limiting their ability to support deeper exploration.

In this paper, we propose a faceted search system built on top of an SKG that enables unit-aware access to measured data extracted from scientific literature. Our system automatically identifies, normalizes, and semantically aligns measured quantities with their respective units, allowing users to filter and compare results even when they originate from different studies using heterogeneous units. By transforming unstructured numerical data into structured facets, the system supports more flexible and meaningful interactions with scholarly content. Additionally, the framework allows for integration with external or remote knowledge graphs, enabling the dynamic enrichment of search results and broader exploration across distributed digital library infrastructures. This contributes to a more intelligent and semantically enriched retrieval environment that enhances the usability and accessibility of scientific knowledge.

As part of our future work, we plan to expand the system’s query capabilities through query expansion techniques, including the automatic recognition of abbreviations, synonyms, and variant expressions of terms. This will increase recall and improve user experience, especially in interdisciplinary research contexts. Furthermore, we aim to integrate systematic review functionalities into the knowledge graph infrastructure, helping researchers identify patterns, synthesize findings, and conduct comprehensive literature analyses. Overall, our work advances the role of digital libraries as intelligent platforms not only for retrieving documents but also for navigating structured, quantitative scientific knowledge. By addressing the challenges of measured data harmonization and semantic search, we contribute toward more open, transparent, and accessible scholarly communication.

\begin{credits}

\subsubsection{\ackname}  This work was co-funded by the European Research Council for the project ScienceGRAPH (Grant agreement ID: 819536) and the TIB Leibniz Information Centre for Science and Technology. The authors would like to thank Hamed Babaei Giglou for helpful comments.

\subsubsection{\discintname}
The authors have no competing interests to declare that are relevant to the content of this article.
\end{credits}

\bibliographystyle{splncs04}
\bibliography{sample-base}

\end{document}